\documentclass{article}
\usepackage{amssymb}
\usepackage{amsmath}
\usepackage{graphicx}
\usepackage{epsfig}
\usepackage{bm}
\usepackage{amssymb}
\usepackage{float}
\usepackage{subfigure}
\usepackage{dcolumn}
\usepackage{cancel}
\usepackage[colorlinks]{hyperref}
\usepackage[usenames,dvipsnames]{color}

\begin{document}

\title{{\LARGE Cosmological aspects of $f(R,T)$ gravity in a simple model
with a parametrization of $q$}}
\author{{\large Ritika Nagpal}$^{1}$ {\large and S. K. J. Pacif}$^{2}$ \\
$^{1}$Department of Mathematics, Vivekananda College,\\
Delhi University, New Delhi 110032, Delhi,\ India \\
$^{2}$Centre for Cosmology and Science Popularization (CCSP),\\
SGT University, Gurugram 122505, Haryana, India \\
Email: $^{1}$ ritikanagpal.math@gmail.com , $^{2}$shibesh.math@gmail.com}
\date{}
\maketitle

\begin{abstract}
In this paper, we have considered a quadratic variation of the deceleration
parameter ($q$) as a function of cosmic time ($t$) which describes a smooth
transition from the decelerating phase of the Universe to an accelerating
one and also show some distinctive feature from the standard model. The
logical move of this article is against the behavior of the future Universe, 
\textit{i.e.} whether the Universe expands forever or ends with a Big Rip,
and we observe that the outcome of the considered parametrization comes in
favor of Big Rip future of the Universe. The whole set up of the
parametrization and solution is taken in $f(R,T)$ theory of gravity for a
spatially flat Friedmann-Lema\^{\i}tre-Robertson-Walker (FLRW) geometry.
Furthermore, we have considered the functional form of $f(R,T)$ function as $%
f(R)+f(T)$, where a quadratic correction of the geometric term $R$ is
adopted as the function $f(R)$, and a linear matter term $f(T)$. We have
investigated some features of the model by examining the behavior of
physical parameters. Our primary goal here is to discuss the physical
dynamics of the model in $f(R,T)$ gravity. We have found, the EoS parameter
also has the same singularity as that of the Hubble parameter \textit{i.e.}
at the initial phase and at the Big Rip. The EoS parameter is explored in
some detail for our choice of $f(R,T)$ function considered here. Different
cases for $f(R,T)$ functional form for different values of the coupling
parameters are discussed, and the evolution of the physical parameters is
shown graphically.
\end{abstract}

\textbf{Keywords:} $f(R,T)$ gravity, Parametrization, Dark Energy, Late-time
acceleration.

\section{Introduction}

The cosmological data of type Ia supernova (SNIa) independently analyzed by
Perlmutter et al. \cite{perlmutter}, and Riess et al. \cite{riess} in
different projects have individually disclosed that the Universe is
undergoing a phase of cosmic acceleration at present. This late-time
acceleration of the Universe is not known exactly, but the inclusion of an
extra source in the energy budget can explain the idea of cosmic
acceleration well. This additional source is assumed to be weird and has an
anti-gravitational effect with highly negative pressure is generally dubbed
as dark energy (DE), \cite{samiDE}, \cite{bamba}, \cite{kamion}, \cite%
{liwang}, \cite{shaifu}. The recent results on SNIa research have shown that
almost $70\%$ of the energy budget is DE. Though, the right candidate for
dark energy is still a point of heavy discussion. Recently, gravitational
wave detection and the picture of black hole shadow strengthen Einstein's
general theory of relativity, and any modifications in Einstein's theory
(specifically to the geometry part) is not much appropriate. However,
Einstein himself was not convinced with the matter distribution in the
Universe \textit{i.e.} the right-hand side of his field equations
(representing matter sector) is considered to be made up of low-grade wood
while the geometry part is of solid marble (representing the space-time).
Any extra source term such as Einstein's cosmological constant (representing
energy density of vacuum) could be added into the energy-momentum tensor and
serve as a candidate for dark energy. The most favored candidate of dark
energy is the well-known cosmological constant $\Lambda $. Also, $\Lambda
CDM $ models have the best fit with many observational datasets. However,
with this significant $\Lambda $, and due to its non-dynamical character
produces a plethora of DE models with the dynamical equation of state
explaining the accelerated expansion in the past few decades such as
quintessence \cite{zlatev}, \cite{brax}, \cite{barreiro}, \cite{albrecht}, 
\cite{quintsami}, \cite{quintsami2}, K-essence \cite{mukhanov1}, \cite%
{mukhanov2}, \cite{chiba}, \cite{steinhardt1}, \cite{steinhardt2}, \cite%
{liddle}, tachyons \cite{sen1}, \cite{sen2}, \cite{garousi1}, \cite{garousi3}%
, \cite{bergshoeff}, \cite{kutasov}, quintom \cite{bof}, \cite{quintom1},
phantom \cite{cladwell}, \cite{phantom1}, \cite{phantom2}. Although dark
energy is not the only possibility, there is another way to explain the
late-time acceleration and is modify the gravity \cite{noji-odi}. So far, a
wide range of modifications in the geometry part of the Einstein field
equation has been done. These include $f(R)$ gravity \cite{farao},
scalar-tensor theories \cite{alimo}, braneworld models \cite{nan} etc. 
\textbf{Many remarkable studies have already been carried out in the field
of alternative gravity theory \cite{StephenE}, \cite{ElizaldeE}, \cite%
{OdintsovE1}, \cite{OdintsovE2}, \cite{OikonomouE1}, \cite{OikonomouE2}, 
\cite{CapozzielloE1}, \cite{LucaE1}, \cite{NojiriE1}, \cite{LucaE2}.}

We have comprehended that modified gravity theories became the main field of
study of modern cosmology, especially due to the motivational search to
explore the elusive nature of DE and the reason of late times cosmic
acceleration \cite{perlmutter}, \cite{riess}. These theories demand the
modification or generalization in the Einstein-Hilbert action and offer
gravitational field equations distinct from the field equations of the DE
model of cosmic speed up \cite{samiDE}, \cite{Nojiri2017a}. There are
numerous ways to depart from GR, so, therefore, researchers developed
several realistic modified gravity theories whose enticing features are
recorded in \cite{Nojiri2007}, \cite{Clifton2012}, \cite{Capozziello2008}, 
\cite{Guo2005}. One of the most simple generalization of $f(R)$ theory of
gravity \cite{Cardone2003}, \cite{Francaviglia2008}, \cite{Laurentis2011}, 
\cite{Nojiri2011}, \cite{Boehmer2008}, \cite{Lobo2009}, \cite{Cardone2006}, 
\cite{Abdalla2005}, \cite{Felice2010}, \textbf{\cite{Sotiriou2010E}, \cite%
{Clifton2011E}, \cite{cappo1E}} was first proposed in $1984$ by Goenner \cite%
{Goenner1984}. The consequence of non-minimal coupling between scalar
curvature and matter Lagrangian was broadly examined in \cite{Nojiriabc}, 
\cite{Harko2010}. Including the above theory, Harko and Lobo extend the
Einstein-Hilbert action by expecting an arbitrary coupling of scalar
curvature and matter Lagrangian. The straightforward exercise of this
arbitrary coupling was initially performed by Poplawski \cite{Poplawski2006}
on the basis of the principle of least action. The outcome of this arbitrary
matter-geometry coupling leads to the violation of conservation of EMT,
which results in the appearance of non-geodesic motion of the massive
particle. In $2011$, Harko \textit{et al.} \cite{Harkofrt} widespread the
sphere of this arbitrary coupling of scalar curvature $R$ and trace $T$ of
EMT, and this introduces another modification in the GR known as $f(R,T)$
theory of gravity. A lot of notable works have already been carried out in $%
f(R,T)$ gravity \cite{Reddy2012}, \cite{Alves2016}, \cite{Moraes2017a}, \cite%
{Yousaf2017}, \cite{Das2016}, \cite{jam}, \cite{Shabani2017a}, \cite%
{Shabani2018a}, \cite{Shabani2018b}, \cite{rtk1}, \cite{rtk2}, \cite{rtk3}, 
\cite{rtk4}, \cite{rtk5}, \cite{rtk6} .\newline

It is to be noted that cosmic acceleration is a late-time phenomenon, and
the structure formation in the Universe during the matter-dominated era
requires a phase of decelerated expansion where gravity must be the
dominating force. So, to illustrate the whole evolution of the Universe, one
needs a phase of super acceleration in the beginning (inflation) and middle
deceleration, and an accelerated phase at late times. This phenomenon is
attributed to the cosmological parameter known as the deceleration
parameter, and the simplest way to obtain such a scenario is the
cosmological parametrization \cite{SKJP2020}, \cite{SKJP2016}. In
literature, there are various schemes of parametrization discussed,
suggesting an early deceleration to the present accelerating era together
with a cosmological phase transition. Various DE models, as well as modified
gravity models, have been explored in the past few years. We are interested
here to discuss a simple model of the Universe that describes the
observational scenario of the different phases of the Universe in $f(R,T)$
gravity with a parametrization of the deceleration parameter that also helps
to find an exact solution of the field equations and try to see the role of
the coupling parameters in the quadratic form of $f(R,T)$ function
considered here. \newline

The paper is organized in the discussed sequences: The first section is the
introduction and describes the present scenario in cosmology. In the second
sect., we have discussed the basic formalism of the $f(R,T)$ gravity. The
field equations are derived in the third sect. for a homogeneous and
isotropic FLRW space-time in the backdrop of $f(R,T)$ gravity. The
parametrization scheme and the solution of field equations are discussed in
the fourth sect. In the fifth sect., the dynamical behavior of the equation
of state parameter is explored, while some special cases for the functional
form of the $f(R,T)$ function are discussed in the sixth sect. Finally, we
have summarized the physical insights of the results in the seventh sect.

\section{Basic formalism in $f(R,T)$ gravity}

The general action of $f(R,T)=f(R)+f(T)$ gravity \cite{Harkofrt} 
\begin{equation}
S=\int \{\frac{1}{16\pi G}f(R,T)+L_{m}\}\sqrt{-g}dx^{4},  \label{1}
\end{equation}

Here, we allow the general form of $f(R,T)$ as $f(R)+f(T)$, where a
quadratic correction of the geometric term $R$ is adopted as the function $%
f(R)=R+\alpha R^{2}$ \cite{star1}, and a linear matter term $f(T)$. The term 
$R^{2}$ turn up in the general functional form of $f(R)$ indicates the
simple corrections to GR. Numerous observations of the Universe are
compatible with the Starobinsky model \cite{pla}, \cite{jdb}, therefore we
have extended the form of $f(R,T)=R+f(T)$ to $f(R,T)=f(R)+f(T)$, so that
acceleration in the Universe (early and late time acceleration) can be
explained by the theories beyond GR \cite{paul}. Also, to establish the
exotic imperfect fluids and taking quantum effects as well in the above
mentioned function $f(R)$, we outset a trace $T$ dependent term which is
responsible for matter Lagrangian $L_{m}$ and exhibit the set of field
equations. Here, in this research, we take $f(T)$ as a first degree function
of trace $T$ defined as $f(T)=2\lambda T$, $\lambda $ being a coupling
constant. Following from this the concluded form of $f(R,T)$ function is $%
R+\alpha R^{2}+2\lambda T$.

Energy momentum tensor(EMT) of matter \cite{Harkofrt} is expressed as 
\begin{equation}  \label{2}
T_{ij}= -\frac{2}{\sqrt{-g}} \frac{\delta(\sqrt{-g}L_m)}{\delta g^{ij}},
\end{equation}
where $T=g^{ij}T_{ij}$ is the trace of EMT. Furthermore, the dependence of $%
L_m$ is dependent on $g_{ij}$, therefore 
\begin{equation}  \label{3}
T_{ij}= g_{ij}L_m-2\frac{\delta L_m}{\delta g^{ij}}.
\end{equation}
Varying the action (\ref{1}) \textit{w.r.t.} $g_{ij}$, we have 
\begin{equation}  \label{4}
f^{R}(R,T) R_{ij}-\frac{1}{2} g_{ij} f(R,T)+(g_{ij}\Box-\nabla_i
\nabla_j)f^{R}(R,T)=8\pi G\, T_{ij}-f^{T}(R,T)(T_{ij}+\Theta_{ij}),
\end{equation}
where $f^{R}(R,T)$ and $f^{T}(R,T)$ act as the derivative of $f(R,T)$ 
\textit{w.r.t.} $R$ and $T$ respectively, $\Box = g^{ij}\nabla_{i}
\nabla_{j} $ represents the d'Alembert operator, where $\nabla_i$ shows the
covariant derivative \textit{w.r.t.} $g_{ij}$. \newline
On defining $\Theta_{ij}$, we have 
\begin{equation}  \label{5}
\Theta_{ij}\equiv g^{lm} \frac{\delta T_{lm}}{\delta g^{ij}}=
-2T_{ij}+g_{ij}S_m-2 g^{lm} \frac{\delta^2 L_m}{\delta g_{ij} \delta g^{lm}}.
\end{equation}
In present paper, we assume perfect fluid matter in the universe, so one can
take the form of matter Lagrangian $L_m = -p$. 
\begin{equation}  \label{6a}
T_{ij}=(\rho +p)u_{i}u_{j}-p g_{ij},
\end{equation}%
where $\rho$ is the energy density and $p $ is the pressure of the fluid
present in the Universe. Equation (\ref{5}) defines the variation of EMT as 
\begin{equation}  \label{7}
\Theta_{ij}=-2T_{ij}-p g_{ij}.
\end{equation}
The gravitational field equations can be obtained as using Eq. (\ref{7}) in
Eq. (\ref{4}) 
\begin{equation}  \label{8}
f^{R}(R,T) R_{ij}-\frac{1}{2}g_{ij} f(R,T)+(g_{ij}\Box-\nabla_i \nabla_j)
f^{R}(R,T)= 8\pi G T_{ij}+f^{T}(R,T)(T_{ij}+p g_{ij}).
\end{equation}

On contracting the above equation (\ref{8}) \textit{w.r.t} $g^{ij}$ and
reorganize the terms in Eq. (\ref{8}), one can read the next two equations
as 
\begin{equation}
Rf^{R}(R,T)-2f(R,T)+3\Box f^{R}(R,T)=8\pi G\,T+(T+4p)f^{T}(R,T).  \label{9}
\end{equation}%
and, we have 
\begin{equation}
R_{ij}=\frac{1}{f^{R}(R,T)}\Big(8\pi G\,T_{ij}+\frac{1}{2}g_{ij}+(\nabla
_{i}\nabla _{j}-g_{ij}\Box )f^{R}(R,T)+f^{T}(R,T)(T_{ij}+pg_{ij})\Big).
\label{10}
\end{equation}%
Next, we define a new operator $\circleddash $, 
\begin{equation}
\circleddash =\nabla _{i}\nabla _{j}-g_{ij}\Box .  \label{11}
\end{equation}%
Using Eq. (\ref{11}) in (\ref{10}), 
\begin{equation}
R_{ij}=\frac{1}{f^{R}(R,T)}\Big(8\pi G\,T_{ij}+\frac{1}{2}
g_{ij}+\circleddash _{ij}f^{R}(R,T)+f^{T}(R,T)(T_{ij}+pg_{ij})\Big).
\label{12}
\end{equation}%
On rewriting the Ricci scalar $R$ in Eq. (\ref{9}), we obtain 
\begin{equation}
R=\frac{1}{f^{R}(R,T)}\Big(8\pi G\,T+2f(R,T)-3\Box
f^{R}(R,T)+(T+4p)f^{T}(R,T)\Big).  \label{13}
\end{equation}

The field equations with the Einstein tensor $G_{ij}$ on the LHS can be
achieved by applying Eqs. (\ref{12}) and (\ref{13}) in Eq. (\ref{8}) 
\begin{eqnarray}
G_{ij}=R_{ij}-\frac{1}{2}Rg_{ij} &=&\frac{8\pi G\,T_{ij}}{f^{R}(R,T)}+\frac{%
1 }{f^{R}(R,T)}\Big[\frac{1}{2}g_{ij}(f(R,T)-Rf^{R}(R,T))+\circleddash
_{ij}f^{R}(R,T)+(T_{ij}+pg_{ij})f^{T}(R,T)\Big],  \label{14} \\
&=&\frac{8\pi G}{f^{R}(R,T)}(T_{ij}+T_{ij}^{^{\prime }}),  \notag
\end{eqnarray}%
where $T_{ij}^{^{\prime }}=\frac{1}{8\pi G}\Big(\frac{1}{2}%
g_{ij}(f(R,T)-Rf^{R}(R,T))+\circleddash
_{ij}f^{R}(R,T)+(T_{ij}+pg_{ij})f^{T}(R,T)\Big).$ From the above field
equations, EFE in GR can be resumed by fixing $\alpha =0$ and $\lambda =0$.
Imposing the Bianchi identity on Eq. (\ref{14}) leads to\footnote{
Note that this equation has been obtained in \cite{Shabani2018a}. However, 
because of the metric signature in the present work, the last term in Eq. ( %
\ref{145}) has obtained the opposite sign.} 
\begin{equation}
\Big(8\pi G+f^{T}(R,T)\Big)\nabla ^{i}T_{ij}+\frac{1}{2}f^{T}(R,T)\nabla
_{i}T+T_{ij}\nabla ^{i}f^{T}(R,T)+\nabla _{j}\Big(pf^{T}(R,T)\Big)=0.
\label{145}
\end{equation}


\section{Field equations in RW geometry}

\qquad In modern physical cosmology, the spatial distribution of the matter
in the Universe is based on cosmological principle, according to which, the
Universe is homogeneous and isotropic on a large scale; therefore, it has no
irregularities on the large scale structure over the course of evolution,
that was primarily identified by big-bang. In addition to that, present
cosmic observations evidence a flat geometry of the Universe; therefore we
assume spatially flat FLRW metric of the Universe of the form 
\begin{equation}
ds^{2}=dt^{2}-a^{2}(t)(dx^{2}+dy^{2}+dz^{2}),  \label{15}
\end{equation}%
where $a(t)$ is the scale factor. The scalar curvature $R$ and trace $T$ of
EMT (\ref{6a}) are defined as

\begin{equation}
R=-6(2H^{2}+\dot{H}),  \label{18}
\end{equation}%
where $H=\frac{\dot{a}}{a}$ represents the Hubble parameter and an overhead
dot shows the derivative \textit{w.r.t.} to $t$. 
\begin{equation}
T=\rho -3p,  \label{17}
\end{equation}%
By considering the form of $f(R,T)=R+\alpha R^{2}+2\lambda T$ and using Eqs.
(\ref{6a}), (\ref{17}), (\ref{18}) in Eq. (\ref{14}), the field equations
read as 
\begin{equation}
\Big[\frac{1}{1+2\alpha R}\Big]3H^{2}=8\pi \rho +\lambda (3\rho -p)+2\alpha
\xi (H,\dot{H},\ddot{H}),  \label{19}
\end{equation}%
\begin{equation}
\Big[\frac{1}{1+2\alpha R}\Big]2\dot{H}+3H^{2}=-8\pi p+\lambda (\rho
-3p)+2\alpha \eta (H,\dot{H},\ddot{H},\dddot{H}),  \label{20}
\end{equation}%
where $\xi (H,\dot{H},\ddot{H})=-9\big(4H^{4}+\dot{H}^{2}-4H^{2}\dot{H}-2H%
\ddot{H}\big)$ and $\eta (H,\dot{H},\ddot{H},\dddot{H})=3\big(-12H^{4}+5\dot{%
	H}^{2}+4H^{2}\dot{H}+12H\ddot{H}+2\dddot{H}\big)$ are the functions of
Hubble parameter $H$ and its derivatives up to third order respectively.
Also, we have set the units so that $G=1$.\newline

Applying the functional form of $f(R,T)$ in Eq. (\ref{145}), which leads to, 
\begin{equation}
\frac{8\pi +3\lambda }{8\pi +2\lambda }\dot{\rho}-\frac{\lambda }{8\pi
+2\lambda }\dot{p}+3H(\rho +p)=0.  \label{201}
\end{equation}%
Eq. (\ref{19}) reads as 
\begin{equation}
\left( 8\pi +3\lambda \right) \rho -\lambda p=\vartheta ,  \label{201-1}
\end{equation}%
where we have defined,%
\begin{equation}
\vartheta \equiv 3H^{2}+18\alpha \left( \dot{H}^{2}-6H^{2}\dot{H}-2H\ddot{H}
\right) .  \label{201-2}
\end{equation}%
Simplifying Eqs. (\ref{201}) and (\ref{201-1}) with the help of (\ref{201-2}%
), we obtain 
\begin{align}
& \rho =\frac{3\vartheta -\lambda \left( 8\pi +2\lambda \right) ^{-1}H^{-1} 
\dot{\vartheta}}{12\left( 2\pi +\lambda \right) },  \label{201-3} \\
& p=\frac{\left( 8\pi +3\lambda \right) \rho -\vartheta }{\lambda }.
\label{201-4}
\end{align}

The above expressions of the density and pressure in equations (\ref{201-3})
and (\ref{201-4}) consist of terms of Hubble parameter $H$ and its
derivatives. This will provide an exact solution to the field equations with
any simple parametrization schemes \cite{SKJP2020}, \cite{SKJP2016}. There
exist various schemes of parametrization available in the literature, and it
is generally referred to as a model-independent way to explore various dark
energy models. For a huge list of various parametrization schemes, one can
refer to the reference \cite{SKJP2020}, \cite{SKJP2016}. Here, in this
paper, we consider a simple algebraic form of the deceleration parameter $q$
and study the physical properties of the Universe in $f(R,T)$ gravity to
examine the role of the $f(R,T)$ coupling parameters $\alpha $ and $\lambda $
in the evolution of the Universe.

\section{Parametrization of $q$ and Solution}

An adhoc choice of generalized time-dependent deceleration parameter $q(t)$
of second degree is discussed in the paper \cite{bakry}, where the
deceleration parameter is considered as, 
\begin{equation}
q(t)=(8\gamma ^{2}-1)-12\gamma t+3t^{2},  \label{21}
\end{equation}%
where $\gamma >0$ is an arbitrary constant. For this parametrization of $%
q(t) $, our model entirely accelerates when $\frac{3t}{4}-\frac{1}{4}\sqrt{
2+3t^{2}}<\gamma <\frac{3t}{4}+\frac{1}{4}\sqrt{2+3t^{2}}$ and decelerates
according as $\gamma <\frac{3t}{4}-\frac{1}{4}\sqrt{2+3t^{2}}$ or $\gamma >%
\frac{3t}{4}+\frac{1}{4}\sqrt{2+3t^{2}}$ and it predicts phase transitions
when $q=0$ at $t=2\gamma \pm \sqrt{\frac{4\gamma ^{2}+1}{3}}$. As it is well
acknowledged that the Universe experiences an accelerating phase in late
time, so the Universe must passed through a phase of slow expansion in the
past \cite{perlmutter}, \cite{riess} in such case the parametrization of
deceleration parameter is rational.\newline
\qquad Using the relation of deceleration parameter with Hubble parameter $%
H(t)$ \cite{berry2017}, Eq. (\ref{21}) yields the explicit form of the
Hubble parameter and the scale factor as, 
\begin{equation}
H(t)=\frac{1}{t(2\gamma -t)(4\gamma -t)},  \label{22}
\end{equation}%
and 
\begin{equation}
a(t)=\beta \frac{\lbrack t(4\gamma -t)]^{\frac{1}{8\gamma ^{2}}}}{(2\gamma
-t)^{\frac{1}{4\gamma ^{2}}}}.  \label{23}
\end{equation}%
where $\beta $ is constant of integration. The explicit form of the
expressions for energy density and pressure can be written using the
equation (\ref{22}) in equations (\ref{201-3}) and (\ref{201-4}).

Now, we have the complete solution to our field equations with explicit
forms of all the geometrical and physical parameters to study the dynamics
of the obtained model. Bakry and Shafeek \cite{bakry} have discussed the
geometrical behavior of the model obtained with this parametrization of $%
q(t) $ in general relativity, wherein they have explored the possibility of $%
\gamma =0.5$ in the redshift range $-1<z<4$. The analyses show the model
have Bang to Rip evolution similar to the one discussed by by Caldwell et
al. \cite{caldwell}.\textbf{\ Some notable works have also been studied to
discuss the finite time singularity \cite{TsujikawaE1},\cite{OikonomouE3}, 
\cite{BahamondeaE1}.}

Our presented work is an extended work of the same, where we want to explore
the physical dynamics of the Universe in $f(R,T)$\ theory of gravity with a
quadratic correction term of the $f(R,T)$ function i.e. $f(R,T)=R+\alpha
R^{2}+2\lambda T$ with two parameters $\alpha $ and $\lambda $ and want to
see the role of the correction terms in the evolution of the physical
parameters. As, we are more interested to discuss the present phase of the
Universe, we formulate these kinematic parameters given in Eqs. (\ref{21})
and (\ref{22}) in terms of redshift $z$ by using the relation of scale
factor as $a(t)=1/(1+z)$ (with the normalizing condition $a_{0}=1$, $a_{0}$
being the value of the scale factor at $t=t_{0}$). Now, the expressions for $%
q$ and $H$ are demonstrated in terms of redshift $z$ as, 
\begin{equation}
H(z)=\frac{H_{0}\left( (\beta (z+1))^{8\gamma ^{2}}+1\right) ^{3/2}}{\left(
\beta ^{8\gamma ^{2}}+1\right) ^{3/2}(z+1)^{4\gamma ^{2}}},  \label{24}
\end{equation}%
\begin{equation}
q(z)=-\frac{(\beta (z+1))^{8\gamma ^{2}}+\gamma ^{2}\left( 4-8(\beta
(z+1))^{8\gamma ^{2}}\right) +1}{(\beta (z+1))^{8\gamma ^{2}}+1}.  \label{25}
\end{equation}

For different values of the model parameter $\gamma $, the evolution of the
deceleration parameter $q$ \textit{w.r.t.} redshift $z$ can be plotted as
follows, 
\begin{figure}[tbp]
\begin{center}
$ 
\begin{array}{c@{\hspace{0.1in}}cc}
\includegraphics[width=3.0 in, height=2.3 in]{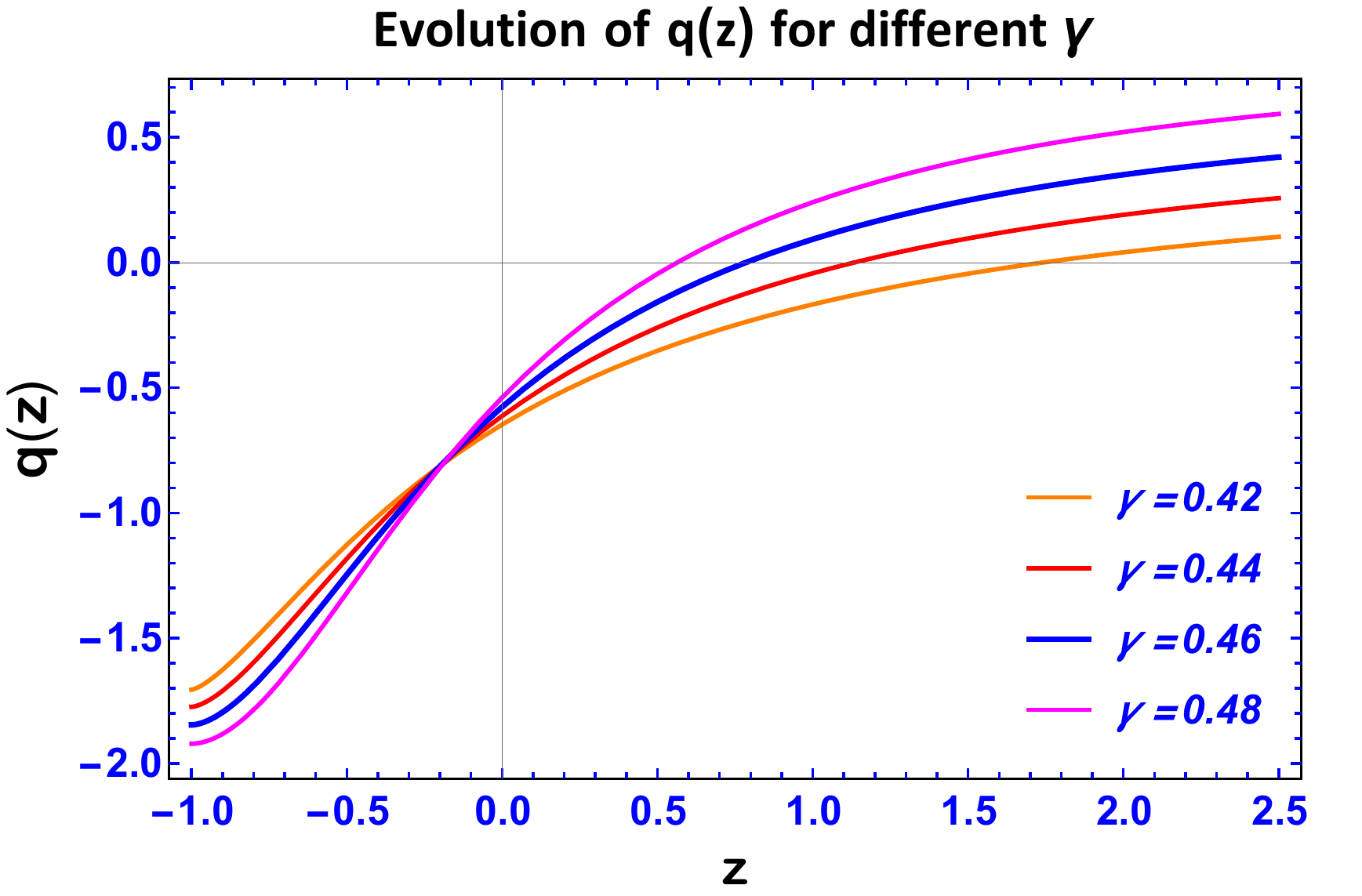} &  &  \\ 
&  & 
\end{array}
$%
\end{center}
\caption{{\protect\scriptsize (a) The plot in the left pannel shows the
evolution of deceleration parameter $q$ \textit{vs.} cosmic time $t$ with
suitable units of time and (b) The right pannel shows the evolution of the
deceleration parameter $q$ \textit{vs.} redshift $z$ for different values of
the model parameter $\protect\gamma $ and a fixed $\protect\beta =1$.}}
\end{figure}

From the above figure, Fig.1 of deceleration parameter, we can interpret the
cosmological phase transition from early deceleration to present
acceleration with the present value of the deceleration parameter $q_{0}<0$
and in the far future, the highly negative value of $q$ indicates the super
acceleration phase leading to a Big Rip singularity.

Pacif et al. \cite{simran} have discussed a model in general relativity with
this same parametrization of $q$ and constrained the model parameters with
some external datasets (Hubble datasets, Supernovae datasets, and Baryonic
Acoustic Oscillation datasets) involved in the model \textit{i.e.}, $\gamma $
and $\beta $ and they have found the values of these model parameters as $%
\gamma =0.44_{-0.01}^{+0.01}$ and $\beta =1.09_{-0.11}^{+0.11}$ for which $%
q_{0}=-0.50_{-0.11}^{+0.12}$ and the phase transition occurs at $%
z_{t}=0.67_{-0.36}^{+0.26}$. Since our discussed model is based on the same
parametrization scheme $q=(8\gamma ^{2}-1)-12\gamma t+3t^{2}$, we can use
these constrained values of the model parameters for our subsequent analysis.

\section{EoS Parameter \& its dynamical behavior}

\label{B}

One of the vital parameter of cosmology is EoS parameter which explains the
different cosmic regimes in Universe. In a more generic way, $\omega $ can
be defined as $p/\rho $ whose diverge values in different ranges discusses
the different cosmic matter in the Universe and the same can be observe in
the table below. 
\begin{table}[tbp]
\caption{Diversity of substances in the Universe}
\label{tabparm}
\par
\begin{center}
\begin{tabular}{lccr}
\hline\hline
&  &  &  \\ 
{\textbf{Substance}} & \,\,\,\,\, \textbf{EoS parameter} \,\,\, & \,\, \, 
\textbf{Observations} &  \\ 
&  &  &  \\ \hline
&  &  &  \\ 
Phantom Universe & $\omega<-1$ & \,\,\,\,\, Repulse Weak Energy Condition
(WEC) Lead to Big Rip &  \\ 
&  &  &  \\ 
Cosmological Constant & $\omega=-1$ & Phantom Universe (Inconsistent with
observations) &  \\ 
&  &  &  \\ 
Quintessence & $\frac{-1}{3}< \omega <-1$ & Cosmological Constant (68\% of
the Universe) &  \\ 
&  &  &  \\ 
Cold matter & $\omega=0$ & Pressurless matter (32\% of the Universe) &  \\ 
&  &  &  \\ 
Hot matter & $0 < \omega <\frac{1}{3}$ & Radiation (Insubstantial at present
time) &  \\ 
&  &  &  \\ 
Radiation & $\omega=\frac{1}{3}$ & Hard Universe (Significant in Early
Universe) &  \\ 
&  &  &  \\ 
Hard Universe & $\frac{1}{3}< \omega <1$ & Exorbitant high densities &  \\ 
&  &  &  \\ 
Stiff matter Universe & $\omega=1$ & Ekpyrotic matter &  \\ 
&  &  &  \\ 
Ekpyrotic matter & $\omega>1$ & Repulse Dominant Energy Condition
(DEC)(Quintessence) &  \\ 
&  &  &  \\ \hline\hline
\end{tabular}
\end{center}
\end{table}

The cosmic acceleration can be attained with the inequality $1+3\omega<0$ as
advised by the Friedmann equations, which can be realized for an exotic
matter related to the negative pressure. The existence of different
substances in the Universe give rise to various cosmic phases. These
different epochs can be observed by varying the EoS parameter $\omega$ (see
Table I).\newline

It is worth considering the behavior of EoS parameter in $f(R,T)$ gravity.
According to our model in $f(R,T)$ gravity, we have realized that in the
initial times, $\omega =\frac{\frac{\left( 8\pi +3\lambda \right) \rho
-\vartheta }{\lambda }}{\frac{3\vartheta -\lambda \left( 8\pi +2\lambda
\right) ^{-1}H^{-1}\dot{\vartheta}}{3\left( 8\pi +4\lambda \right) }}$ turns
to an expression of coupling constant $\lambda $ (of matter and geometry)
and the model parameter $\gamma $ but remains unrelated of the other
coupling parameter $\alpha $. In the late times, it can be realized that the
EoS parameter $\omega $ depends solely on $f(R,T)$ coupling constant $%
\lambda $.

Mathematically, we can write, 
\begin{equation}
\omega _{i}\equiv \lim_{t\rightarrow 0}\omega =\frac{4\pi (-3+32\gamma
^{2})+3(-1+16\gamma ^{2})}{12\pi +(3+16\gamma ^{2})\lambda },  \label{w1}
\end{equation}

and 
\begin{equation}
\omega _{f}\equiv \lim_{t\rightarrow \infty }\omega =-3+\frac{8\pi }{\lambda 
}.  \label{w2}
\end{equation}

Here, the subscript `$i$' indicates the value of EoS parameter in the early
Universe and the subscript `$f$' represents the value of EoS parameter in
the late Universe. We can notice that the expression of $\omega _{i}$ is not
tied-up with the parameter $\alpha $ and $\omega _{f}$ is free from $\alpha $
and $\gamma $.

Some elementary calculations witness that constraining the EoS parameter $%
\omega$ as $0<\omega_{i}<1$ (which is plausible in cosmology) forces one to
prefer

\begin{itemize}
\item[(i)] $\lambda >0$\thinspace\ \thinspace and\thinspace \thinspace\ $ 
\frac{1}{4}\sqrt{3}\sqrt{\frac{(4\pi +\lambda )}{(8\pi +3\lambda )}}<\gamma
< \sqrt{\frac{3}{4}}$, \newline
and the positivity of $\frac{(4\pi +\lambda )}{(8\pi +3\lambda )}$ further 
leads the condition on the $f(R,T)$ coupling constant $\lambda <-4\pi $ 
\thinspace \thinspace\ or \thinspace \thinspace\ $\lambda >-\frac{8\pi }{3}$
, \newline

\item[(ii)] The EoS parameter can behave like ekpyrotic matter $(\omega >1)$
by choosing $\lambda >0$ \thinspace \thinspace\ and $\gamma >\sqrt{\frac{3}{%
4 }}$ (for positive coupling $f(R,T)$ constant $\lambda $), and \newline

\item[(iii)] (a) $\lambda < -4 \pi$ and \,\, $\gamma > \sqrt{\frac{3}{4}}$ 
\,\, or\,\, (b) $-4 \pi < \lambda < -2 \pi$ \,\, and \,\, $\frac{1}{4}  
\sqrt {3} \sqrt{-\frac{4\pi+\lambda}{\lambda}} < \gamma < \sqrt{\frac{3}{4}}$
\,\,or\,\, (c) $-2 \pi < \lambda < 0$ \,\, and \,\, $\sqrt{\frac{3}{4}} < 
\gamma < \frac{1}{4} \sqrt{3} \sqrt{\frac{-4\pi + \lambda} {\lambda}}$
\end{itemize}

which resists dominant energy condition (DEC) followed by matter-dominated
era where energy density is extensively high to the radiation-dominated era
in the initial Universe. Some fundamental facts about different eras of
cosmic evolution can be comprehended by the results (\ref{w1}) and (\ref{w2}%
). Also, from (\ref{w1}), one can arbitrate the fundamental gravitational
model which portray two different cosmic phase transitions:

\begin{itemize}
\item[(i)] for $\gamma =\frac{\sqrt{3}}{4}$, we get $\omega _{i}^{s}=1$. It 
is simply observed that the stiff matter epoch of the Universe can be 
obtained by setting $\gamma =\frac{\sqrt{3}}{4}$ and remain independent of 
other model parameters. In this case, the model explains a transition from a
state in which stiff matter dominates in the early eras to a state which 
behaves like the DE in the late times.

\item[(ii)] Also, it is feasible to obtain the following relation on model 
parameter $\gamma$ that shows a transition in the initial Universe $
\omega^{p}_{i}=0 $ (a pressure-less matter dominated era) to a Universe with
DE in the late times.  
\begin{align}  \label{28}
\gamma = \frac{\sqrt{3}}{4} \frac{\sqrt{4 \pi + \lambda}}{\sqrt{8 \pi + 3
\lambda}}.
\end{align}

\item[(iii)] From Eq. (\ref{w2}) one more analysis can be performed on EoS 
parameter $\omega $. As we all know that, the standard $\Lambda $CDM model 
fits well with various measurements and cosmic observational data \cite{abd}
, \cite{hin} and is based on the most consistent and prevailing big bang 
scenario. In accord with many observations \cite{abd}, \cite{hin}, $\Lambda $
CDM model is considered as one of the candidates to describe the accelerated
expansion in the Universe. Therefore, it is worthwhile to discuss the 
conditions from (\ref{w2}), in our model, which acts like cosmological 
constant $\omega =-1$ in the late time. The negative coupling between matter
and geometry \textit{i.e.} $\lambda =-2\pi $ is required to get the stated 
condition. Again, it is require to remark that the behavior of our model 
with standard $\Lambda $CDM depends only on the model constant $\lambda $.
\end{itemize}

From the above computations, it has been inferred that the parameter $\alpha 
$ does not play any role in deciding the initial and final stages of cosmic
evolution, which simply imply that the only changes in the matter part of
the Lagrangian $\lambda $ is responsible for different states of the cosmic
evolution. Moreover, it is always worthwhile to study the behavior of matter
density and pressure in the limit large times. Straightforward calculations
reveal that both terms $\rho $ and $p$ tend to $0$ provided that the model
parameter $\gamma >0$ and is free from the model constants $\lambda $ and $%
\alpha $. Mathematically, the limit of large times gives 
\begin{equation}
\lim_{t\rightarrow \infty }\rho =\lim_{t\rightarrow \infty }p=0.  \label{29}
\end{equation}

In the next section, we shall discuss the behavior of the physical
parameters \textit{e.g.} energy density, pressure and EoS parameter in $%
f(R,T)$ gravity with the quadratic form of the $f(R,T)$ function with the
two coupling parameters $\alpha $ and $\lambda $. The following special
cases arise for which we can discuss the physical evolution of the energy
density, pressure, and also the EoS parameter through some graphical
representations.

\section{Special cases for $f(R,T)$ function}

The following four cases arise for different values of the coupling
parameters are showing the role of the correction terms in $f(R,T)$ gravity.

\subsection{case I: $\protect\lambda =0$, $\protect\alpha =0$ i.e. $f(R,T)=R$%
}

For these vanishing values of the coupling parameters ($\lambda =0$ \& $%
\alpha =0$), the case reduces to standard general relativity. The evolution
of the physical parameters for this case is described in some detail in the
ref. \cite{bakry}.

%
%

To understand the recent past, present, and future evolution of the physical
parameters $\rho $, $p$ and $\omega $, we plot them with respect to redshift 
$z$, which are shown in the following graphical representations in the
following figure, Fig.2.

\begin{figure}[tbp]
\begin{center}
$ 
\begin{array}{c@{\hspace{0.1in}}cc}
\includegraphics[width=2.3 in, height=2.3 in]{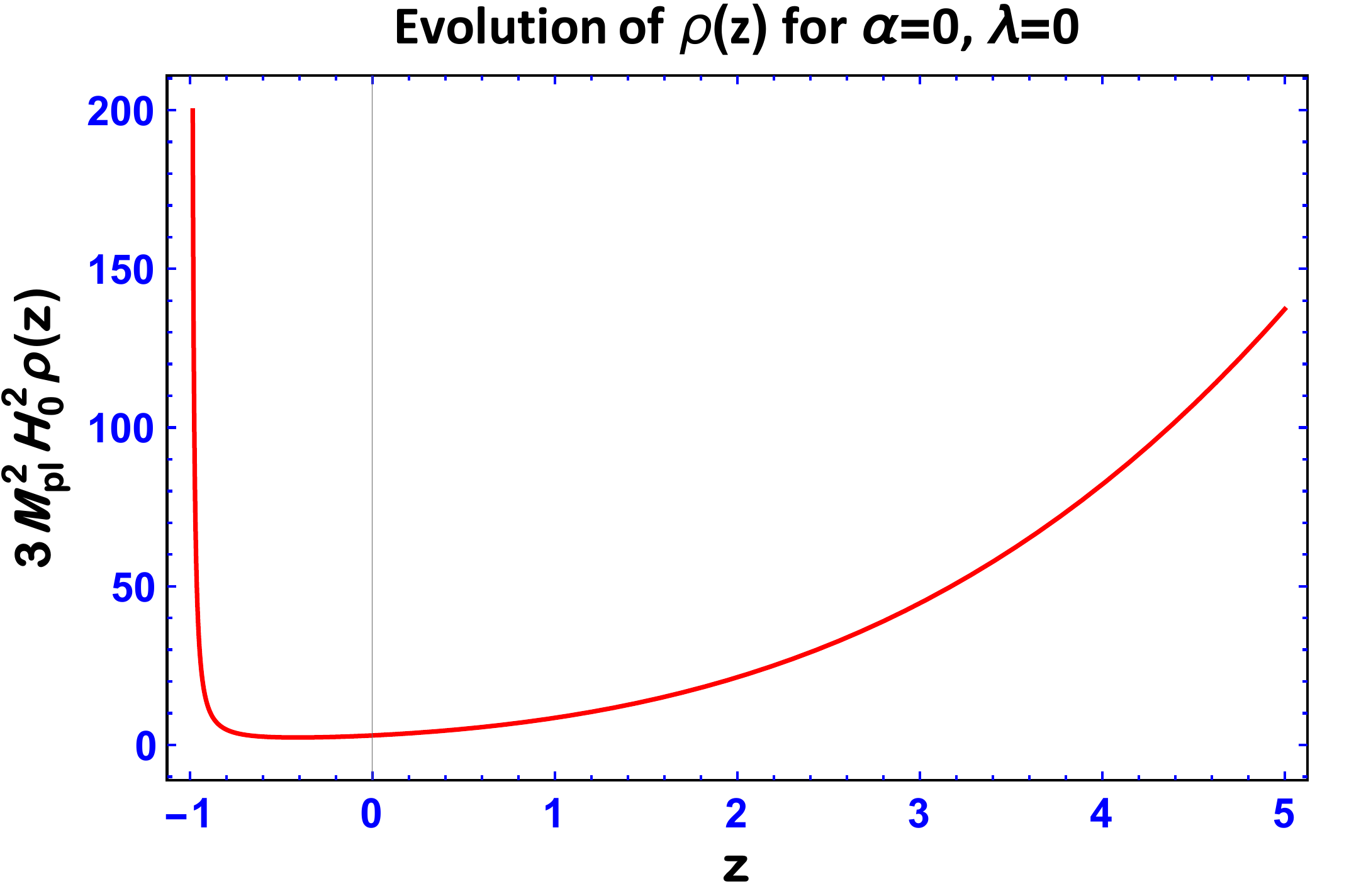} & %
\includegraphics[width=2.3 in, height=2.3 in]{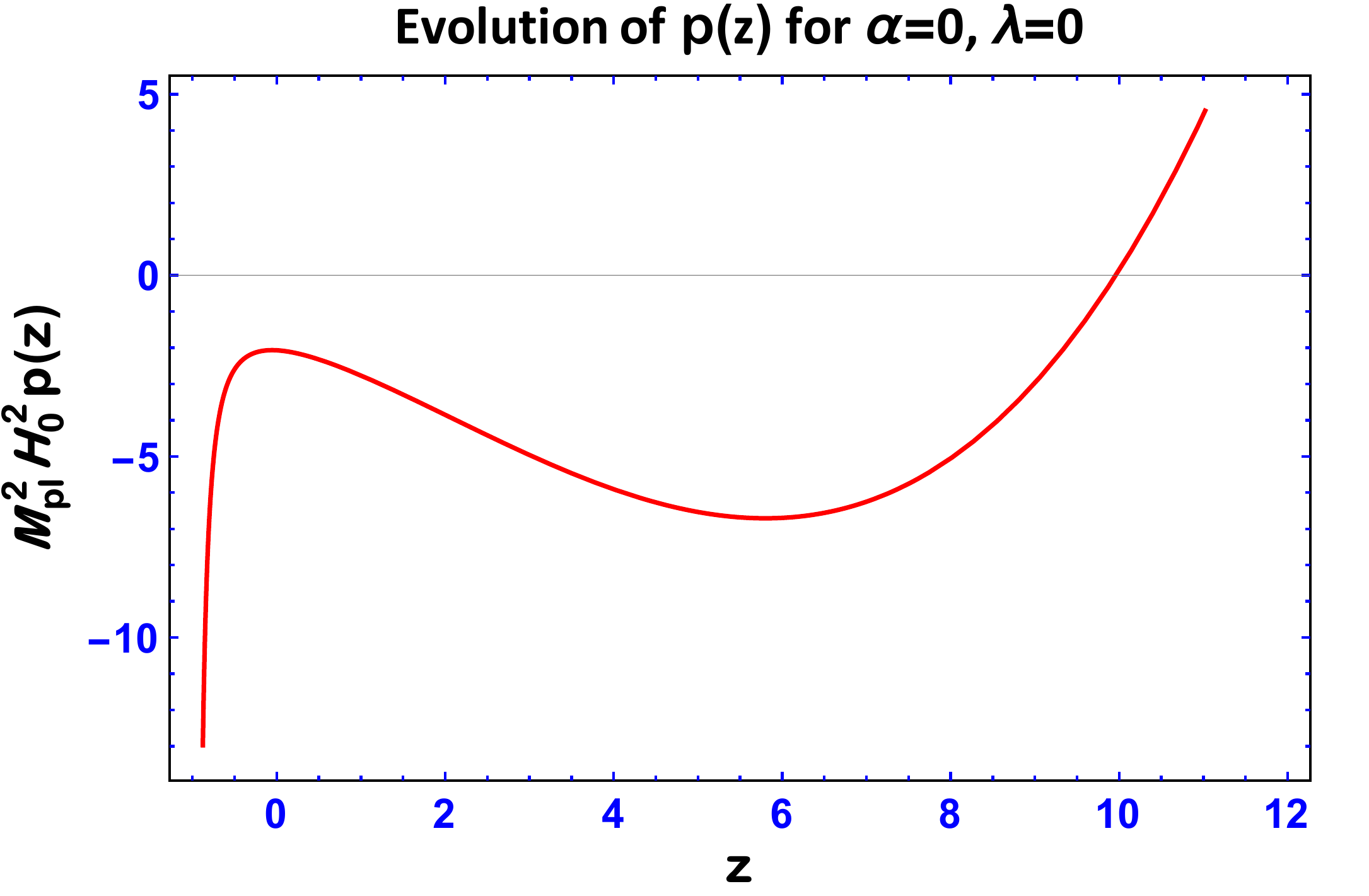} & %
\includegraphics[width=2.3 in, height=2.3 in]{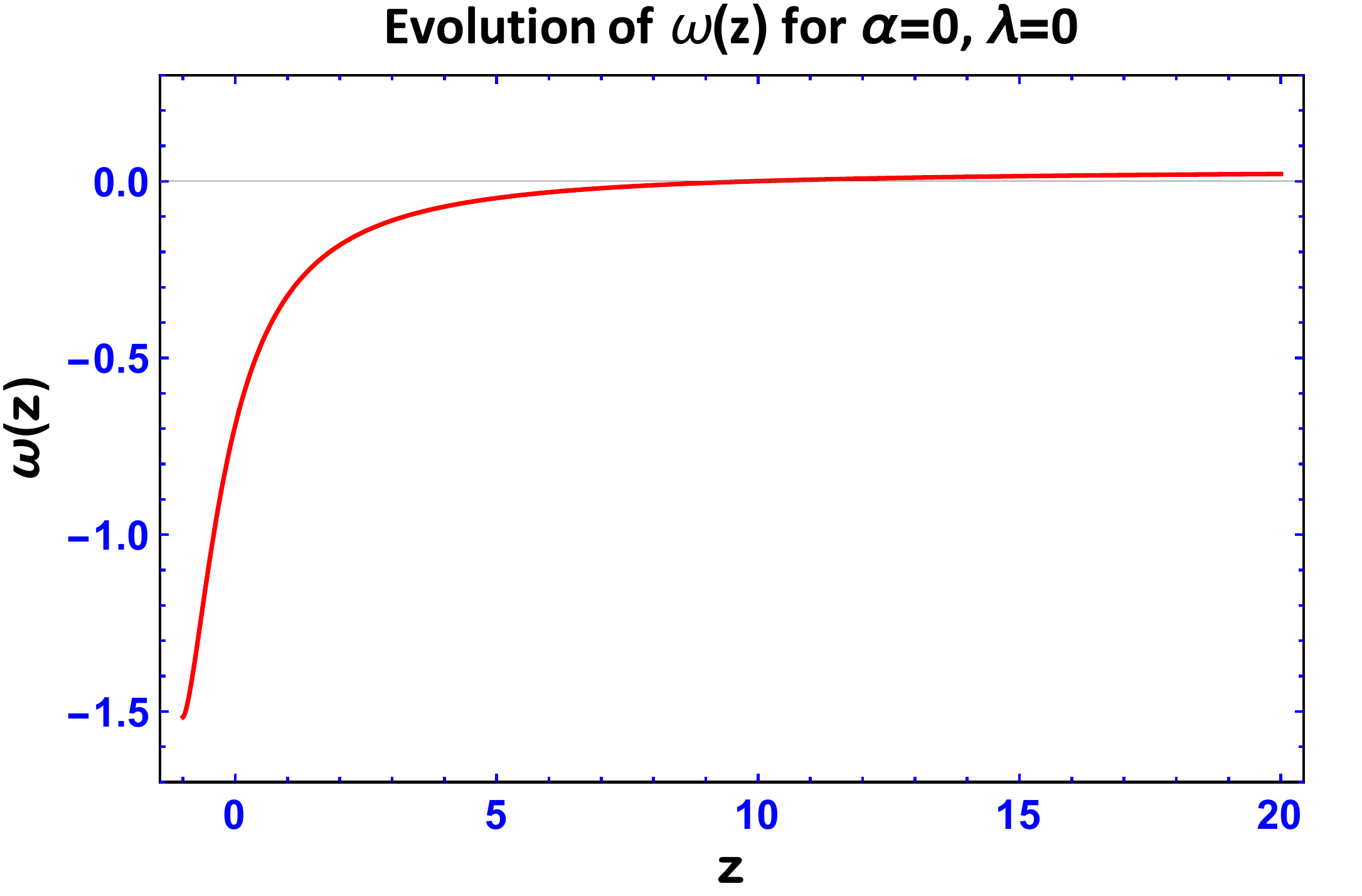} \\ 
\mbox (a) & \mbox (b) & \mbox (c)%
\end{array}
$%
\end{center}
\caption{{\protect\scriptsize (a) The plot of energy density $\protect\rho %
\sim z$, (b) The plot of cosmic pressure $p\sim z$ and (c) The plot of EoS $%
\protect\omega \sim z$ for $\protect\gamma =0.44$ and $\protect\beta =1.09$.
In this plot the energy scale is} $(8\protect\pi G)^{-1/2}=M_{pl}$ ($G=1$
here).}
\end{figure}

\subsection{case II: $\protect\alpha =0$, $\protect\lambda \neq 0$ i.e. $%
f(R,T)=R+2\protect\lambda T$}

%
%

For this case also, to understand the recent past, present, and future
evolution of the physical parameters $\rho $, $p$ and $\omega $, we plot
them with respect to redshift $z$, which are shown in the following
graphical representations in the following figure, Fig.3.

\begin{figure}[tbp]
\begin{center}
$ 
\begin{array}{c@{\hspace{0.1in}}cc}
\includegraphics[width=2.3 in, height=2.3 in]{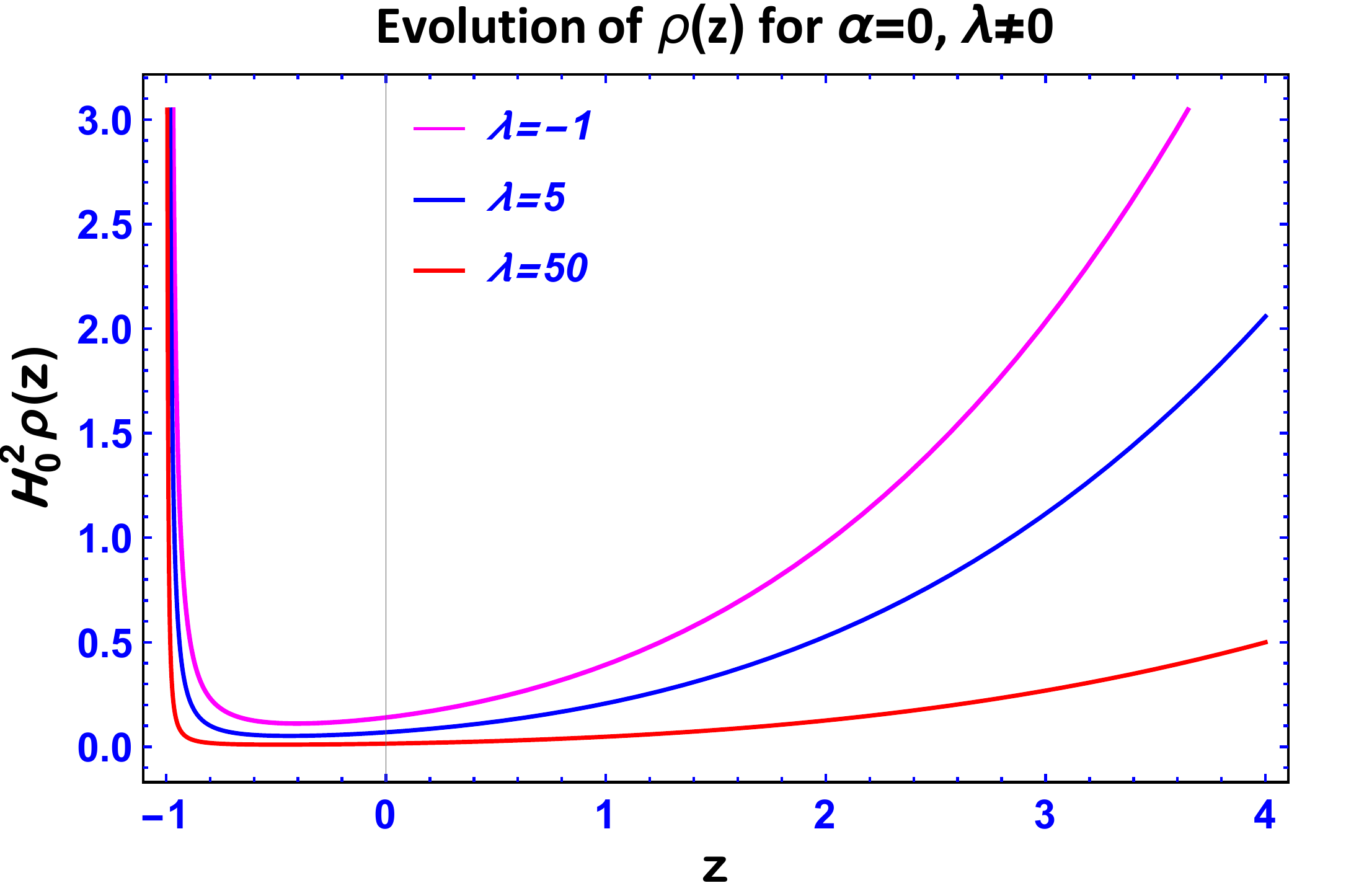} & %
\includegraphics[width=2.3 in, height=2.3 in]{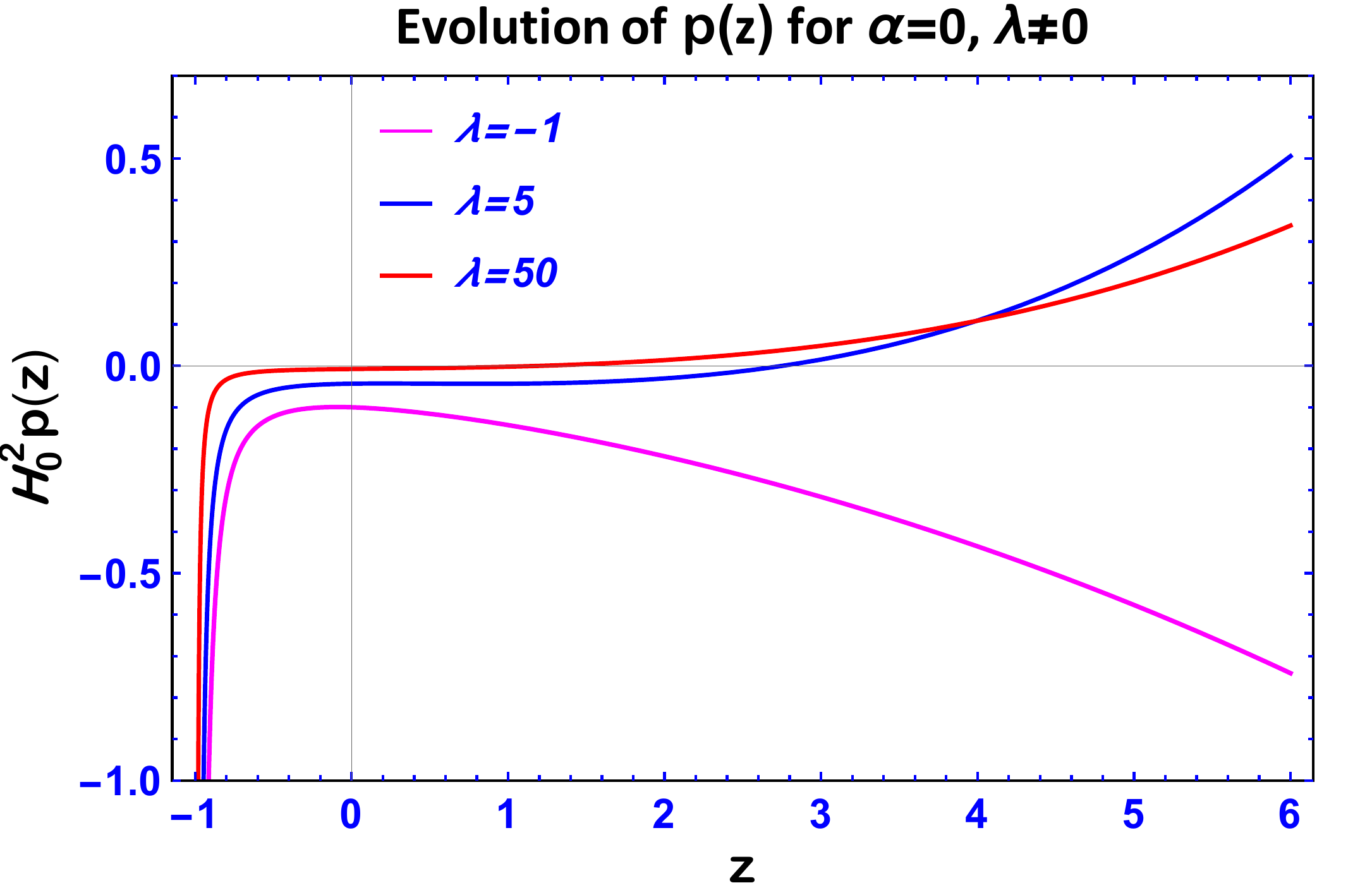} & %
\includegraphics[width=2.3 in, height=2.3 in]{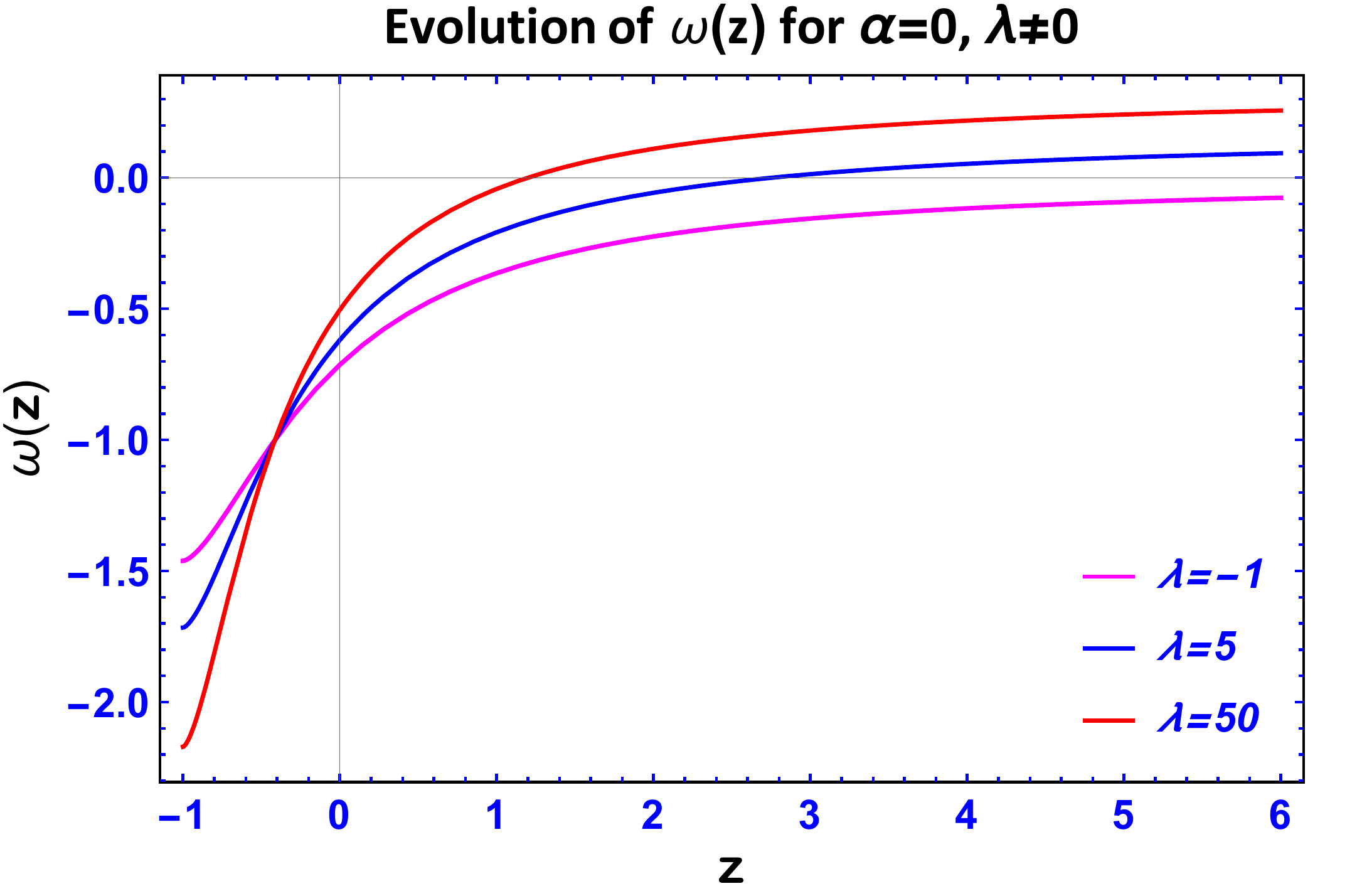} \\ 
\mbox (a) & \mbox (b) & \mbox (c)%
\end{array}
$%
\end{center}
\caption{{\protect\scriptsize (a) The plot of energy density $\protect\rho %
\sim z$, (b) The plot of cosmic pressure $p\sim z$ and (c) The plot of EoS $%
\protect\omega \sim z$ for $\protect\gamma =0.44$ and $\protect\beta =1.09$.}
}
\end{figure}

From the above plotted Fig. 3, we can interpret that the negative value of
the coupling parameter $\lambda $ is incompatible with the present scenario
of the late-time Universe as expected. Furthermore, it is found that other
two cases for the values of the coupling parameters with $\alpha \neq 0$, $%
\lambda =0$ i.e. $f(R,T)=R+\alpha R^{2}$ and $\alpha \neq 0$, $\lambda \neq
0 $ i.e. $f(R,T)=R+\alpha R^{2}+2\lambda T$ are incompatible with the
considered parametrization of $q$.

\section{Discussion and conclusion}

In this work, we have examined the cosmological model in the framework of
FLRW space-time using the non-linear alternative theory of gravity, namely $%
f(R,T)$ gravity. The dynamics of the model using the specific form of $%
f(R,T)=R+\alpha R^{2}+2\lambda T$ is investigated in section (\ref{A}).
Thus, the behavior of the Universe is based on the preferred choice of $%
f(R,T)$ function, which pulls out an explicit set of field equations.
Additionally, this article is an attempt to design a cosmic model by taking
a suitable parametrization of the deceleration parameter, which was first
considered in the paper \cite{bakry} wherein two phenomena Big Rip and Big
Bang of the Universe were discussed together with the cosmic evolution in
the general theory of relativity. Later on, the same model is studied in the
paper \cite{simran} where the authors found some observational constraints
using some external datasets and also discussed the statefinder diagnostics.
Here, in our study, we have taken the motivation from \cite{bakry} and \cite%
{simran} and extended the study in $f(R,T)$ gravity wherein our main
intention is to study the physical parameters (especially EoS $\omega $) in $%
f(R,T)$ gravity with quadratic correction terms i.e. $f(R,T)=R+\alpha
R^{2}+2\lambda T$. Some distinctive features of the model are recorded and
discussed as follows.

\begin{itemize}
\item The parametrization of second-degree time-dependent deceleration 
parameter $q(t)$ has been chosen such that cosmos passes through different 
phases depending on the value of model parameter $\gamma $. The functional 
form of $q(t)$ exhibits different regimes of the Universe as the model 
demonstrates the bouncing criteria depending on the values of $\gamma $. The
model begins with Big bang at time $t=0$ and ends at $4\gamma $ with $
q=8\gamma ^{2}-1$ in both scenarios. The model completes one cycle in the 
time range $t\in (0,4\gamma )$, \textit{i.e.} Universe is in the stage of 
Big Rip at $t=2\gamma $ while elapses through $t\in (0,2\gamma )$ and 
recurring at the stage of Big Bang at $t=4\gamma $ while passes through $
t\in (0,2\gamma )$. The flipping behavior of the Universe in two different 
time ranges is the remarkable feature of the parametrization (\ref{21}).

\item As we have mentioned that the motive of our study is to examine the 
dynamics of physical parameters in the framework of $f(R,T)$ gravity, 
thereby it is worthwhile to understand the working of EoS parameter $\omega $
. In accord with our findings, we have noted that our cosmic model in the 
early times remains unaffected by the quadratic correction term $\alpha $, 
wherein the effect of $\lambda $ and model parameter $\gamma $ is 
significant. Furthermore, $\omega $ is only $\lambda $ dependent in the late
time. Consequently, the role of $\gamma $, $\lambda $, and $\alpha $ make 
our results utterly different from the findings of \cite{bakry}.

\item A comprehensive analysis of $\omega $ has been performed in section ( %
\ref{B}) with the view to understand the existence of various substances and
their dynamic behavior in the Universe, which is classified as EoS parameter
(see Table I).

\item The analyses in section (\ref{B}) demonstrates that the parameter $
\alpha $ does not play any vital role in deciding the initial and final 
stages of cosmic evolution. The only variations in the matter part of the 
Lagrangian coupling constant $\lambda $ are responsible for various states 
of the cosmological evolution. Straightforward calculations reveal that both
terms $\rho $ and $p$ tend to $0$ (\ref{29}) provided that the model 
parameter $\gamma >0$ and is free from the model constants $\lambda $ and $
\alpha $.

\item To see the evolution of the physical parameters energy density ($\rho $
), pressure ($p$) and equation of state parameter ($\omega $) in the recent 
past, present and future evolution, we have plotted them \textit{w.r.t.} the
redshift ($z$) for two cases $\alpha =0$, $\lambda =0$ (case-I corresponding
to GR) and $\alpha =0$, $\lambda \neq 0$ (case-II). The other two cases $
\alpha \neq 0$, $\lambda =0$ and $\alpha \neq 0$, $\lambda \neq 0$ are 
incompatible with the discussed scenario.
\end{itemize}

\indent By inspecting all the above points, one can interpret that this
cosmological model describes a cyclic Universe scenario with the considered
scheme of parametrization of deceleration parameter and reconstructing some
physical parameters in $f(R,T)$ theory of gravity. The above study imparts a
reason to understand several cosmic scenarios right from the evolution of
the Universe (Big Bang) to its end (Big Rip). Undoubtedly, the integration
of observational cosmology in this study provides a more precise range to
model parameters so that the behavior of geometrical and physical parameters
could be investigated in a more suitable way. However, the existing study is
only an attempt to figure out the dynamics of the physical parameters of the
Universe.

\end{document}